# Warp and blur model for imaging through turbulence.

**MIKHAIL CHARNOTSKII**
*Erie. CO, 80516, USA*
*Mikhail.Charnotskii@gmail.com*

**Abstract:** Point Spread Function (PSF) for imaging through inhomogeneous refractive medium, such as atmospheric turbulence is bounded by three constraints [Charnotskii, Opt. Eng., **52**, 04600, (2013)]. PSF is non-negative, band-limited, and the third constraint, related to the energy conservation principle, warrants the absence of fluctuations in the image of a uniformly bright object. We develop a version of the common warp and blur model for the anisoplanatic image distortions by turbulence that satisfies these three constraints.

1. Introduction

Investigation of the image distortions for imaging through atmospheric turbulence has been an active research subject for more than 60 years, [1 – 3]. Currently, statistics of turbulent distortions is well understood, [2, 4, 5], and research is concentrated on the modeling and mitigation of the turbulent distortions, [6-13]. Typically, imaging models involve generation of the random point spread function (PSF) that combines the effects of both optical system and turbulence and convolution of this PSF with the original undisturbed object or scene. For the majority of practical turbulence imaging scenarios, imaging is anisoplanatic, and PSF is a function of two two-dimensional arguments: position vectors in the object and image planes.

It is common in the image modeling to reduce the infinite variety of the turbulent distortions to just two random components: geometrical distortions, represented as a random shift of the turbulent PSF or warp, and residual random PSF spread, or blur, [6 – 13]. Accordingly, the turbulence mitigation techniques employ dewarping, [6, 8, 10, 11, 12], and deblurring, [10 - 12, 14, 15] approaches.

In order to realistically represent the turbulent distortions, any PSF has to satisfy certain functional constrains [16]. The focus of the present investigation is development of a simple Warp and Blur (W&B) of the turbulent imaging model that is compliant with the functional constraints on the turbulent PSF.

The paper is organized as follows: Section 2 presents the three functional constraints on the turbulent PSF. Section 3 formulates the W&B model for anisoplanatic turbulent PSF, and shows how the W&P model can be made conforming to the functional constraints of the previous section. Section 4 proposes a specific example of the S&B model that is based on the simple, anisoplanatic phase screen propagation concept. It also discusses the warp and blur statistics related to the phase screen propagator. Conclusions summarizes the findings and formulates suggestions for the further efforts

2. Three constraints on a PSF

Irradiance distribution $I(\mathbf{R}_I)$ in the image plane for the incoherent object $O(\mathbf{R}_O)$ can always be presented as

$$I(\mathbf{R}_I) = \iint d^2 R_O O(\mathbf{R}_O) PSF(\mathbf{R}_I - \mathbf{R}_O, \mathbf{R}_O), \tag{1}$$

where $PSF(\mathbf{R}, \mathbf{R}_O)$ is the anisoplanatic Point Spread Function (PSF), and the object plane representation is used for the image. In paraxial approximation PSF can be presented as, [16]

$$\begin{aligned}PSF(\mathbf{R}, \mathbf{R}_O) = \frac{k^2}{4\pi^2 L^2 \Sigma} \iint d^2 R_A \iint d^2 r_A A\left(\mathbf{R}_A + \frac{\mathbf{r}_A}{2}\right) A\left(\mathbf{R}_A - \frac{\mathbf{r}_A}{2}\right) \exp\left(\frac{ik}{L}\mathbf{R} \cdot \mathbf{r}_A\right) \\ \times g\left(\mathbf{R}_O, L; \mathbf{R}_A + \frac{\mathbf{r}_A}{2}, 0\right) g^*\left(\mathbf{R}_O, L; \mathbf{R}_A - \frac{\mathbf{r}_A}{2}, 0\right).\end{aligned} \tag{2}$$

Here $k = 2\pi/\lambda$ is the wavenumber, object plane is $z = L$, aperture $A(\mathbf{R}_A)$ of the imaging system optical system is located at the plane $z = 0$. Factors $g(\mathbf{r}_1, z_1; \mathbf{r}_2, z_2)$ accounts for all turbulence perturbations of a spherical wave propagation from the point $(\mathbf{r}_1, z_1)$ to point $(\mathbf{r}_2, z_2)$. As introduced by Eq. (2), the PSF for the free-space case, $g(\mathbf{r}_1, z_1; \mathbf{r}_2, z_2) = 1$, is normalized to a unit total power.

PSF is the image of the point source located at $(\mathbf{R}_O, L)$, and therefor is **non-negative**. This is the first PSF constraint. Anisoplanatic Optical Transfer Function (OTF) is the Fourier transform of the PSF

$$OTF(\mathbf{\kappa}, \mathbf{R}_O) = \frac{1}{4\pi^2} \iint d^2R \ PSF(\mathbf{R}, \mathbf{R}_O) \exp(-i\mathbf{R} \cdot \mathbf{\kappa})$$
$$= \frac{1}{4\pi^2 \Sigma} \iint d^2R_A A\left(\mathbf{R}_A + \frac{\mathbf{\kappa}L}{2k}\right) A\left(\mathbf{R}_A - \frac{\mathbf{\kappa}L}{2k}\right) g\left(\mathbf{R}, L; \mathbf{R}_A + \frac{\mathbf{\kappa}L}{2k}, 0\right) g^*\left(\mathbf{R}, L; \mathbf{R}_A - \frac{\mathbf{\kappa}L}{2k}, 0\right), \quad (3)$$

where

$$\Sigma = \iint d^2R_A A^2(\mathbf{R}_A) \quad (4)$$

is the aperture area.

Eq. (3) reveals that for any finite "hard edge" aperture OTF is identically zero for spatial frequencies outside some wave vectors domain determined by the aperture shape and size. This brings about the second PDF constraint: PSF is **band-limited** regardless of the turbulence disturbances. This constrained sometimes is ignored in the literature when Gaussian PSFs are used, [12].

The third, less well-known constraint, is valid only for the refractive turbulence case, where total power of propagating wave is conserved, [17]. As was elaborated in [16], the third constraint can be presented as

$$\iint d^2R_O PSF(\mathbf{R} - \mathbf{R}_O, \mathbf{R}_O) = 1. \quad (5)$$

This equality is trivial for the free-space and isoplanatic, phase screen, cases, but it also holds for any random sample of the refractive turbulence PSF. To the best of the author's knowledge this constrained is never invoked in the turbulence imaging publications since it was described in [16] with the sole exception of [18]. As follows from Eq. (5), the third constraint requires that the image of the uniformly bright object is not affected by refractive turbulence. The further implication is that the local object contrast plays the dominant role in formation of the image turbulence fluctuations [16 – 19].

As a reference, for the top hat aperture with radius $a$ diffraction limited PSF and OTF are

$$PSF^{(0)}(\mathbf{R}) = \frac{1}{\pi R^2} J_1^2\left(\frac{kaR}{L}\right),$$
$$OTF^{(0)}(\mathbf{\kappa}) = \frac{1}{2\pi^3}\left[\arccos\left(\frac{\kappa L}{2ka}\right) - \frac{\kappa L}{2ka}\sqrt{1 - \left(\frac{\kappa L}{2ka}\right)^2}\right], \quad (6)$$

and diffraction cutoff is $\kappa^{(0)} = 2ka/L$. This cutoff still persists for the general case of turbulent distortions, Eq. (3).

Several statistics of the turbulent PSF have been discussed in [5]. Specifically, the variances of the Strehl number, $S$, and PSF power, $P$, and their covariance have been calculated.

$$S(\mathbf{R}_O) \equiv \frac{PSF(0, \mathbf{R}_O)}{PSF^{(0)}(0)}, \ P(\mathbf{R}_O) \equiv \iint d^2R \, PSF(\mathbf{R}, \mathbf{R}_O) \quad (7)$$

It was found that for the wide range of imaging conditions the Strehl number is the product of the two statistically-independent factors. The first factor is the power flux through the imaging aperture $P(\mathbf{R}_O)$. The second factor, $p(0, \mathbf{R}_O)$ corresponds to the Strehl number of the PSF normalized to the unit power, and can be attributed to the blur fluctuations. Formally, this can be presented as follows

$$\langle P(\mathbf{R}_O) p(0, \mathbf{R}_O) \rangle = \langle P(\mathbf{R}_O) \rangle \langle p(0, \mathbf{R}_O) \rangle, \ p(\mathbf{R}, \mathbf{R}_O) \equiv \frac{PSF(\mathbf{R}, \mathbf{R}_O)}{P(\mathbf{R}_O)}. \quad (8)$$

Unlike the three strict PSF constraints described earlier, that are mandatory for every PSF sample, Eq. (8) is a statistical feature of the PSF fluctuations.

## 3. Warp and blur imaging model

Turbulent PSF $PSF(\mathbf{R}, \mathbf{R}_O)$ is a non-negative, band-limited nonhomogeneous two-dimensional random field with respect to the variable $\mathbf{R}$, where $\mathbf{R}_O$ is an anisoplanatic parameter. It is common in the literature to reduce functional variations of the PSF for a fixed $\mathbf{R}_O$ to just a few parameters. This reduce the computational burden compare to the direct numeric simulation [13]. In the simplest case the multitude of the PSF variations is reduced to the random anisoplanatic warp and blur. We write a W&B model PSF as

$$PSF(\mathbf{R}, \mathbf{R}_O) = P(\mathbf{R}_O) F(\mathbf{R} - \mathbf{w}(\mathbf{R}_O), b(\mathbf{R}_O)). \tag{9}$$

Here $F(\mathbf{R}, b)$ is deterministic non-negative band-limited PSF normalized to the unit power. Two dimensional vector random field $\mathbf{w}(\mathbf{R}_O)$ describes geometrical distortions, commonly called warp. Parameter $b$ determines the PSF blur, but preserves the band-limited property of $F(\mathbf{R}, b)$, and is introduced in such way that $F(\mathbf{R}, 0)$ is just a diffraction-limited PSF. Obviously, factor $P(\mathbf{R}_O)$ is the random total PSF power. It is important to note that W&P model implicitly assumes that the spatial scales of the warp and blur changes are larger than the PSF width.

Multitude of the turbulence mitigation techniques attempts to improve the image quality by removing the turbulent warp [6, 8, 10, 11]. Note that warp compensation widens the image bandwidth beyond the diffraction cutoff and can result in the turbulence super-resolution, [8, 15, 20]. Further image improvement can be achieved by deconvolution [?], which attempts to reduce the PSF blur. Interestingly, the instantaneous PSF power is never discussed in the frame work of the turbulence image processing.

This can be achieved by a random scaling of diffraction-limited PSF preserving the PSF power, and not exceeding the bandwidth constraint

$$F(\mathbf{R}, b) = (b+1)^2 F(\mathbf{R}(b+1), 0), \quad b \geq 0. \tag{10}$$

This approach would require a model for a scalar random field $b(\mathbf{R}_O)$ with non-Gaussian distribution. More important, however, is that each PSF sample would have diffraction cutoff

$$\kappa(b) = \frac{ka}{(b+1)L}^2 \leq \kappa^{(0)}, \tag{11}$$

and have no power in the $(\kappa(b), \kappa^{(0)})$ spectral range. This does not match the physical picture of the turbulence blur, which is associated with random dephasing of the spectral components, which are still present in the shot-exposure image up to the diffraction limit.

We suggest to relate blur to the phase perturbations at the imaging aperture. Namely, we replace the $g$ factors by the phasors, as follows

$$PSF(\mathbf{R}, \mathbf{R}_O) = \frac{k^2}{4\pi^2 L^2 \Sigma} \iint d^2 R_A \iint d^2 r_A A\left(\mathbf{R}_A + \frac{\mathbf{r}_A}{2}\right) A\left(\mathbf{R}_A - \frac{\mathbf{r}_A}{2}\right) \times \exp\left(\frac{ik}{L} \mathbf{R} \cdot \mathbf{r}_A + i\varphi\left(\mathbf{R}_O, \mathbf{R}_A + \frac{\mathbf{r}_A}{2}\right) - i\varphi\left(\mathbf{R}_O, \mathbf{R}_A - \frac{\mathbf{r}_A}{2}\right)\right). \tag{12}$$

The phase model, Eq. (12) preserves unit PSF power, and essentially applies to the normalized PSF $F(\mathbf{R}, b)$ in Eq. (9). For the S&B model, in order to keep a single blur parameter, we limit phase perturbation to a simple defocus, or quadratic phase,

$$\varphi(\mathbf{R}_O, \mathbf{R}_A) = \frac{k}{2L} b(\mathbf{R}_O) R_A^2, \tag{13}$$

For a top hat aperture normalized PSF $F(\mathbf{R}, b)$ can be presented as

$$F(\mathbf{R}, b) = \frac{N^2}{\pi} \left(\int_0^1 x\, dx \cos(Nbx^2) J_0(x\rho)\right)^2 + \frac{N^2}{\pi} \left(\int_0^1 x\, dx \sin(Nbx^2) J_0(x\rho)\right)^2. \tag{14}$$

Here, $N = ka^2/L$ is the aperture Fresnel number, and $\rho = kaR/L$ is the image plane coordinate normalized by diffraction resolution. Fig. 1 shows examples of the normalized PSF, Eq. (14) as function of $\rho$. Parameter is defocus – Fresnel number product $Nb$.

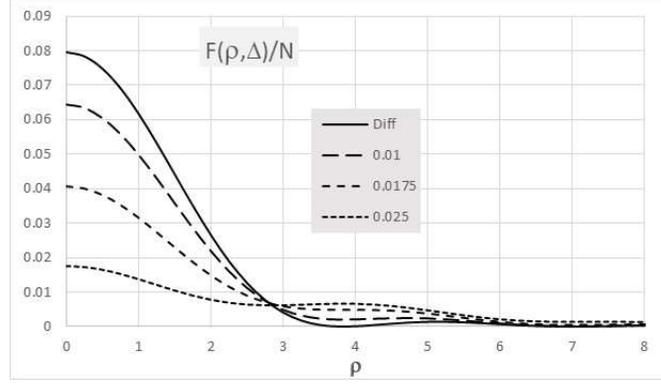

Fig. 1. Normalized PSF, Eq. (14). Parameter is $Nb$.

For W&B PSF the third restrain, Eq. (5) is

$$\iint d^2 R_O P(\mathbf{R}_O) F(\mathbf{R} - \mathbf{R}_O - \mathbf{w}(\mathbf{R}_O), b(\mathbf{R}_O)) = 1, \quad \forall \mathbf{R}. \tag{15}$$

Warp field $\mathbf{w}(\mathbf{R}_O)$ defines a map from the object, $\mathbf{R}_O$, plane to the image, $\mathbf{R}_I$, plane

$$\mathbf{R}_O \Rightarrow \mathbf{R}_I; \quad \mathbf{R}_I(\mathbf{R}_O) = \mathbf{R}_O + \mathbf{w}(\mathbf{R}_O). \tag{16}$$

We assume that the inverse map from the image, $\mathbf{R}_I$, plane to the object, $\mathbf{R}_O$, plane also exists and can be presented as

$$\mathbf{R}_I \Rightarrow \mathbf{R}_O; \quad \mathbf{R}_O(\mathbf{R}_I) = \mathbf{R}_I - \boldsymbol{\omega}(\mathbf{R}_I). \tag{17}$$

Obviously,

$$\mathbf{R}_O(\mathbf{R}_I(\mathbf{R}_O)) = \mathbf{R}_O, \quad \mathbf{R}_I(\mathbf{R}_O(\mathbf{R}_I)) = \mathbf{R}_I, \tag{18}$$

and hence

$$\mathbf{w}(\mathbf{R}_O(\mathbf{R}_I)) = \boldsymbol{\omega}(\mathbf{R}_I), \quad \boldsymbol{\omega}(\mathbf{R}_I(\mathbf{R}_O)) = \mathbf{w}(\mathbf{R}_O). \tag{19}$$

Change of integration variable in Eq. (15) to the image plane leads to

$$\iint d^2 R_I \left| \frac{D(\mathbf{R}_O)}{D(\mathbf{R}_I)} \right| P(\mathbf{R}_O(\mathbf{R}_I)) F(\mathbf{R} - \mathbf{R}_I, b(\mathbf{R}_O(\mathbf{R}_I))) = 1, \quad \forall \mathbf{R}. \tag{20}$$

Here

$$\frac{D(\mathbf{R}_O)}{D(\mathbf{R}_I)} = \begin{pmatrix} 1 - \dfrac{\partial \omega_x}{\partial X_I} & -\dfrac{\partial \omega_y}{\partial X_I} \\ -\dfrac{\partial \omega_x}{\partial Y_I} & 1 - \dfrac{\partial \omega_y}{\partial Y_I} \end{pmatrix} \tag{21}$$

is the Jacobian of transformation from the object to the image plane. Evidently, owing to the unit power normalization of $F(\mathbf{R}, b)$, the third PSF constraint, Eq. (20), is satisfied when

$$P(\mathbf{R}_O(\mathbf{R}_I)) = \left| \frac{D(\mathbf{R}_O)}{D(\mathbf{R}_I)} \right|^{-1}, \tag{22}$$

Eq. (20) is formulated in terms of the image plane coordinate $\mathbf{R}_I$. It is convenient to use the inverse Jacobian

$$\frac{D(\mathbf{R}_I)}{D(\mathbf{R}_O)} = \begin{pmatrix} 1 + \dfrac{\partial w_x}{\partial X_O} & \dfrac{\partial w_y}{\partial X_O} \\ \dfrac{\partial w_x}{\partial Y_O} & 1 + \dfrac{\partial w_y}{\partial Y_O} \end{pmatrix}, \tag{23}$$

and PSF power is presented in terms of the object plane coordinates as

$$P(\mathbf{R}_O) = \left|\frac{D(\mathbf{R}_I)}{D(\mathbf{R}_O)}\right| = \left(1 + \frac{\partial w_x}{\partial X_O}\right)\left(1 + \frac{\partial w_y}{\partial Y_O}\right) - \frac{\partial w_x}{\partial Y_O}\frac{\partial w_y}{\partial X_O} \tag{24}$$

This leads to the main result of this work: S&A PSF conforming to the third constraint necessary has the form

$$PSF(\mathbf{R},\mathbf{R}_O) = \left|\left(1 + \frac{\partial w_x}{\partial X_O}\right)\left(1 + \frac{\partial w_y}{\partial Y_O}\right) - \frac{\partial w_x}{\partial Y_O}\frac{\partial w_y}{\partial X_O}\right| F(\mathbf{R} - \mathbf{w}(\mathbf{R}_O), b(\mathbf{R}_O)). \tag{25}$$

Modeling of the turbulent S&A PSF requires generation of two random fields supported at the object plane – vector warp field $\mathbf{w}(\mathbf{R}_O)$ and scalar blur field $b(\mathbf{R}_O)$. Somewhat surprising is that the PSF power is not functionally related to the blur, but is fully determined by warp. This matches our earlier result, [5] that power fluctuations are uncorrelated with the directivity variations. Here, normalized PSF term is used instead of directivity.

Both warp and blur of the normalized PSF $F(\mathbf{R},b)$ can be represented by a formal Taylor expansion of anisoplanatic phase screen in Eq. (12) consisting only of tip-tilt and defocus components. Namely we set

$$g(\mathbf{R}_O, L; \mathbf{R}_A, 0) = \exp\left[i\varphi(\mathbf{R}_O) + i\nabla\varphi(\mathbf{R}_O)\cdot\mathbf{R}_A + \frac{i}{2}\Delta\varphi(\mathbf{R}_O)R_A^2\right], \tag{26}$$

and warp and blur are correspondingly

$$\mathbf{w}(\mathbf{R}_O) = -\frac{L}{k}\nabla\varphi(\mathbf{R}_O), \quad b(\mathbf{R}_O) = \frac{L}{k}\Delta\varphi(\mathbf{R}_O), \tag{27}$$

## 4. Anisoplanatic phase screen and W&B PSF

### 4.1. Geometrical optics phase

The anisoplanatic phase screen (APS) propagation model, also sometimes called Extended Huygens–Fresnel principle, was discussed in [16, 21], where it was concluded that it inherently flawed. In particular, it violates the energy conservation principle, and therefore does not comply with the third PSF constraint, Eq. (5). However, here we attempt to use this propagation model for the normalized PSF $F(\mathbf{R},b)$ only, while delegating the third constraint compliance to the power factor $P(\mathbf{R}_O)$, as discussed in the previous section.

Here we examine the W&B PSF implementation for a simple approximation when all turbulence effects between the object and aperture planes are reduced to a simple geometrical phase. Specifically, we use the APS approximation of Eq. (12), and present phase of spherical wave in geometrical optics approximation as

$$\varphi(\mathbf{R}_O, \mathbf{R}_A) = k\int_0^L dz\, n\left(\mathbf{R}_O\frac{z}{L} + \mathbf{R}_A\left(1 - \frac{z}{L}\right), z\right). \tag{28}$$

Here $n(\mathbf{R},z)$ is the random field of turbulent refractive index fluctuations. If the warp is introduced as PSF centroid displacement, then it is known, e. g. [4], to be just an aperture-averaged phase gradient

$$\mathbf{w}(\mathbf{R}_O) = -\frac{L}{k}\overline{\nabla\varphi(\mathbf{R}_O)} = -\frac{1}{\Sigma}\int_0^L dz\,(L-z)\iint d^2R_A A^2(\mathbf{R}_A)\nabla n\left(\mathbf{R}_O\frac{z}{L} + \mathbf{R}_A\left(1 - \frac{z}{L}\right)\right). \tag{29}$$

The blur by defocus term in Eq. (27) is determined by the aperture-averaged phase front curvature

$$b(\mathbf{R}_O) = \frac{L}{k}\overline{\Delta\varphi(\mathbf{R}_O)} = \frac{1}{L\Sigma}\int_0^L dz\,(L-z)^2\iint d^2R_A A^2(\mathbf{R}_A)\Delta n\left(\mathbf{R}_O\frac{z}{L} + \mathbf{R}_A\left(1 - \frac{z}{L}\right)\right), \tag{30}$$

According to Eq. (24) the linear in $n(\mathbf{R},z)$ PSF power is calculated from Eq. (29) as

$$P(\mathbf{R}_O) \approx 1 + \frac{\partial w_x}{\partial X_O} + \frac{\partial w_y}{\partial Y_O} = 1 - \frac{1}{\Sigma}\int_0^L dz\,(L-z)\frac{z}{L}\iint d^2R_A A^2(\mathbf{R}_A)\Delta n\left(\mathbf{R}_O\frac{z}{L} + \mathbf{R}_A\left(1 - \frac{z}{L}\right)\right) \tag{31}$$

Notably, both blur, Eq. (30) and power, Eq. (31) are determined by transverse Laplacian of the refractive index suggesting the high correlation level of the blur-power correlation. This apparently contradicts the observations of the previous section, where weak blur-power correlation was recognized. We attribute this to the noted shortcomings of the geometrical approximation phase, Eq. (28).

*4.2 Warp and blur statistics*

Warp and blur statistics can be readily calculated using the conventional Markov approximation, [22],

$$\langle n(\mathbf{R},z)n(\mathbf{R}',z')\rangle = 2\pi \iint d^2\kappa \, \Phi_n(\kappa,0)\exp[i\kappa\cdot(\mathbf{R}-\mathbf{R}')]\delta(z-z'), \tag{32}$$

where $\Phi_n(\kappa,\kappa_z)$ is the 3-D spectrum of refractive index fluctuations.

For isotropic turbulence warp is a zero mean homogeneous isotropic vector field on the object plane. Therefore components of the warp covariance tensor are most effectively presented in terms of the warp component parallel and orthogonal to the separation vector

$$\begin{aligned} C_\|(r_O) &\equiv \left\langle \left(\mathbf{W}(\mathbf{R}_O)\cdot\frac{\mathbf{r}_O}{r_O}\right)\left(\mathbf{W}(\mathbf{R}_O+\mathbf{r}_O)\cdot\frac{\mathbf{r}_O}{r_O}\right)\right\rangle, \\ C_\perp(r_O) &\equiv \left\langle \left(\mathbf{W}(\mathbf{R}_O)\times\frac{\mathbf{r}_O}{r_O}\right)\left(\mathbf{W}(\mathbf{R}_O+\mathbf{r}_O)\times\frac{\mathbf{r}_O}{r_O}\right)\right\rangle \end{aligned} \tag{33}$$

and cross correlation of these components is zero for isotropic fields. For a circular symmetric aperture components of the correlation tensor can be presented as

$$\begin{aligned} C_\|(r_O) &= 2\pi^2 L^2 \int_0^L dz\left(1-\frac{z}{L}\right)^2 \int_0^\infty \kappa^3 d\kappa \, \Phi_n(\kappa)\left|\hat{A}\left(\kappa\left(1-\frac{z}{L}\right)\right)\right|^2 \left[J_0\left(\kappa r_O\frac{z}{L}\right)-J_2\left(\kappa r_O\frac{z}{L}\right)\right], \\ C_\perp(r_O) &= 2\pi^2 L^2 \int_0^L dz\left(1-\frac{z}{L}\right)^2 \int_0^\infty \kappa^3 d\kappa \, \Phi_n(\kappa)\left|\hat{A}\left(\kappa\left(1-\frac{z}{L}\right)\right)\right|^2 \left[J_0\left(\kappa r_O\frac{z}{L}\right)+J_2\left(\kappa r_O\frac{z}{L}\right)\right]. \end{aligned} \tag{34}$$

Here

$$\hat{A}(\kappa) \equiv \frac{1}{\Sigma}\iint d^2R \, A^2(\mathbf{R})\exp(-i\kappa\cdot\mathbf{R}) \tag{35}$$

is the normalized Fourier transform of the aperture function.

Blur is a scalar zero-mean homogeneous random field on the object plane with covariance function

$$D(r_O) \equiv \langle b(\mathbf{R}_O)b(\mathbf{R}_O+\mathbf{r}_O)\rangle = 4\pi^2 L^2 \int_0^L dz\left(1-\frac{z}{L}\right)^4 \int_0^\infty \kappa^5 d\kappa \, \Phi_n(\kappa)\left|\hat{A}\left(\kappa\left(1-\frac{z}{L}\right)\right)\right|^2 J_0\left(\kappa r_O\frac{z}{L}\right). \tag{36}$$

Warp correlates only with the parallel component of the warp, and

$$E(r_O) \equiv \left\langle \left(\mathbf{W}(\mathbf{R}_O)\cdot\frac{\mathbf{r}_O}{r_O}\right) b(\mathbf{R}_O+\mathbf{r}_O)\right\rangle = -4\pi^2 L^2 \int_0^L dz\left(1-\frac{z}{L}\right)^3 \int_0^\infty \kappa^4 d\kappa \, \Phi_n(\kappa)\left|\hat{A}\left(\kappa\left(1-\frac{z}{L}\right)\right)\right|^2 J_1\left(\kappa r_O\frac{z}{L}\right). \tag{37}$$

We illustrate these results for the case of the top hat aperture, when normalized aperture Fourier transform is

$$\hat{A}(\kappa) = \frac{2}{\kappa a}J_1(\kappa a), \tag{38}$$

and power law spectrum

$$\Phi(\kappa) = C_\Phi C_n^2 \kappa^{-3-\gamma}. \tag{39}$$

In this case covariances are presented as

$$C_{\parallel}(r_O) = 8\pi^2 C_\Phi L^3 a^{\gamma-1} \int_0^1 dt\, C_n^2(Lt) \int_0^\infty x^{-2-\gamma} dx\, J_1^2(x(1-t)) \left[ J_0\left(x\frac{r_O}{a}t\right) - J_2\left(x\frac{r_O}{a}t\right) \right],$$

$$C_{\perp}(r_O) = 8\pi^2 C_\Phi L^3 a^{\gamma-1} \int_0^1 dt\, C_n^2(Lt) \int_0^\infty \kappa^{-2-\gamma} d\kappa\, J_1^2(x(1-t)) \left[ J_0\left(x\frac{r_O}{a}t\right) + J_2\left(x\frac{r_O}{a}t\right) \right]. \tag{40}$$

Blur covariance is

$$D(r_O) = 16\pi^2 C_\Phi L^3 a^{\gamma-3} \int_0^1 dt\, C_n^2(Lt)(1-t)^2 \int_0^\infty x^{-\gamma} dx\, J_1^2(x(1-t)) J_0\left(x\frac{r_O}{a}t\right), \tag{41}$$

and warp-blur covariance is

$$E(r_O) = -16\pi^2 C_\Phi L^3 a^{\gamma-2} \int_0^1 dl\, C_n^2(Lt)(1-t) \int_0^\infty x^{-1-\gamma} dx\, J_1^2(x(1-t)) J_1\left(x\frac{r_O}{a}t\right), \tag{42}$$

Blur and warp variances are

$$C_{\parallel}(0) = 8\pi^2 C_\Phi C_W L^2 a^{\gamma-1} \int_0^L dz\, C_n^2(z)\left(1 - \frac{z}{L}\right)^{1+\gamma}, \quad C_W = \frac{\Gamma(2+\gamma)\Gamma\left(\frac{1-\gamma}{2}\right)}{2^{2+\gamma} \Gamma^2\left(\frac{3+\gamma}{2}\right)\Gamma\left(\frac{5+\gamma}{2}\right)},$$

$$D(0) = 16\pi^2 C_\Phi C_B L^2 a^{\gamma-3} \int_0^L dz\, C_n^2(z)\left(1 - \frac{z}{L}\right)^{1+\gamma}, \quad C_B = \frac{\Gamma(\gamma)\Gamma\left(\frac{3-\gamma}{2}\right)}{2^\gamma \Gamma^2\left(\frac{1+\gamma}{2}\right)\Gamma\left(\frac{3+\gamma}{2}\right)}. \tag{43}$$

In the Kolmogorov turbulence case, $\gamma = 2/3$, we have

$$C_\Phi = 0.033 \quad C_W = 0.864 \quad C_B = 0.660, \tag{44}$$

and warp and blur variances for homogeneous turbulence path are

$$C_{\parallel}(0) = 0.844 C_n^2 L^3 a^{-1/3}, \quad D(0) = 1.290 C_n^2 L^3 a^{-7/3}. \tag{45}$$

First equation is the classic V. I. Tatarskii result, [23, Eq. (56.23a)]. As should be expected from the geometrical optics approximation used here, correlation coefficients of warp and blur are scaled by the aperture size $a$. Fig. 2 presents numerically calculated auto-correlation coefficients of warp and blur derived from Eqs. (40, 41), and cross-correlation coefficient of warp and blur, Eq. (42). Notable is the weak correlation between warp and blur. This suggests that warp and blur can be modeled as independent random fields. Also remarkable the slow falloff of the warp autocorrelation with separation compare to the blur autocorrelation. This implies that the blur varies much faster across the field of view than warp. Asymptotic analysis of Eqs. (40, 41, 42), not detailed here, shows that

$$\left.\frac{C_{\parallel}(r_O)}{C_{\parallel}(0)}\right|_{r_O > a} \propto \left(\frac{r_O}{a}\right)^{\gamma-1}, \quad \left.\frac{C_{\perp}(r_O)}{C_{\perp}(0)}\right|_{r_O > a} \propto \left(\frac{r_O}{a}\right)^{\gamma-1}, \quad \left.\frac{D(r_O)}{D(0)}\right|_{r_O > a} \propto \left(\frac{r_O}{a}\right)^{-1}. \tag{46}$$

Corresponding asymptotes are shown as dashed lines in Fig. 2.

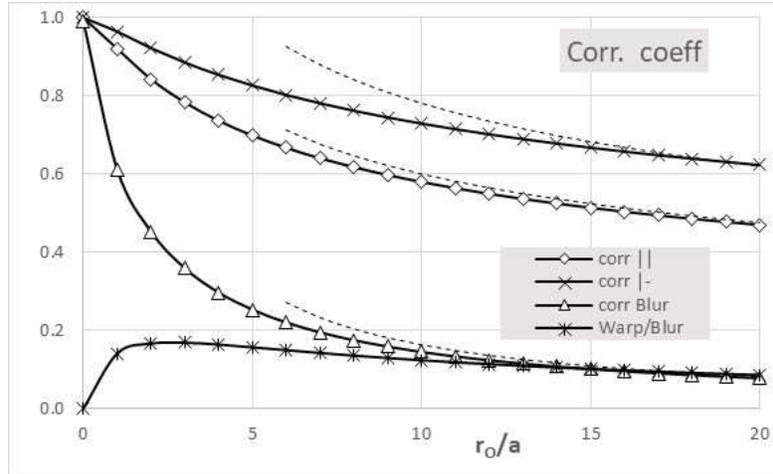

Fig. 2. Correlation coefficients .of warp and blur. Dashed curves – asymptotes for $r > a$ .

## 5. Conclusions

Imaging of incoherent objects through random medium is fully characterized by a random anisoplanatic PSF. In general PSF is non-negative and has a finite bandwidth determined by the aperture size. In case of refractive random medium, such as atmospheric turbulence, there exists a third PSF constraint, presented here by Eq. (5).

Warp and blur PSF model typically reduces random PSF variation to the random shift and spread. In order to maintain the band-limited feature in the presence of the anisoplanatic blur we propose to use anisoplanatic phase perturbations at the aperture plane, which in the simplest case can be a random anisoplanatic defocus, Eq. (13) and Fig. 1.

Compliance of the W&B PSF with the third constraint requires implementation of the additional parameter – random PSF power. We showed that the PSF power is proportional to the Jacobian of the warp map from the object plane to the image plane, Eq. (24). This is the main finding of this work. It implies that PSF power is not functionally related to the blur, but is fully determined by warp map. This matches our earlier results, based on the rigorous statistical analysis of the PSF fluctuations, [5], where it was found that power fluctuations are uncorrelated with the PSF width.

Finally, the general form of the W&B PSF compliant with all three constraints is presented by Eq. (25). Computer simulation of the physically accurate W&B PSFs across the field of view requires generation of a random 2-D isotropic random warp field and a scalar blur field. Both fields are supported at object plane limited by the field of view of the optical system. It is possible that warp and blur are correlated. The PSF power can be calculated from the warp field based on the Eq. (24). Assuming that the commonly used Fourier space based simulation techniques, [24], is used, the power calculations are relatively straightforward. We leave the technical details of these simulations for the further publication.

We illustrated some aspects of our model on a simple example of anisoplanatic phase screen. This imaging mode, with all it known shortcomings, allows the straightforward calculations of the joint warp and blur statistics. We found that the warp correlations declines very slowly across the object plane, while the blur decorrelates much faster. Also, the warp/blur correlation is weak, allowing simulation of warp and blur as independent random fields, thus simplifying the simulation algorithm.